\documentclass[twocolumn,nofootinbib,showpacs,epsfig,prb,reprint]{revtex4-1}

\usepackage{amsmath}
\usepackage{amsfonts}
\usepackage{amssymb}
\usepackage{graphicx}
\usepackage{color}
\usepackage[usenames,dvipsnames]{xcolor}

\usepackage[normal]{subfigure}

\newcommand{\be}{\begin{equation}}
\newcommand{\ee}{\end{equation}}
\newcommand{\reff}[1]{(\ref{#1})}
\newcommand{\rmd}{{\rm{d}}}
\newcommand{\rme}[1]{{\rm{e}}^{#1}}
\newcommand{\mal}{\mathcal}
\newcommand{\dd}[2]{\frac{\rmd^{#1}#2}{\left( 2\pi \right)^{#1 }}}

\newcommand{\ket}[1]{| #1 \rangle}
\newcommand{\bra}[1]{\langle #1 |}
\newcommand{\bracket}[2]{\langle #1 |  #2 \rangle}

\newcommand{\av}[1]{\left\langle #1 \right\rangle}
\newcommand{\ob}{\mathcal{O}}

\newcommand{\lqq}{\left[}
\newcommand{\rqq}{\right]}

\newcommand{\abs}[1]{\left| #1 \right|}

\newcommand{\cosa}[1]{\cos \left(  #1 \right)}
\newcommand{\sina}[1]{\sin \left(  #1 \right)}

\newcommand{\change}[1]{\textcolor{black}{#1}}

\newcommand{\supp}[1]{}

\begin{document}

\title{Response functions after a quantum quench}

\author{Matteo Marcuzzi}
\affiliation{School of Physics and Astronomy, University of Nottingham, Nottingham, NG7 2RD, United Kingdom} 
\affiliation{SISSA --- International School for Advanced Studies, via Bonomea 265, 34136 Trieste, Italy}
\affiliation{INFN --- Istituto Nazionale di Fisica Nucleare, sezione di Trieste}%

\author{Andrea Gambassi}
\affiliation{SISSA --- International School for Advanced Studies, via Bonomea 265, 34136 Trieste, Italy}
\affiliation{INFN --- Istituto Nazionale di Fisica Nucleare, sezione di Trieste}%

\begin{abstract}
The response of physical systems to external perturbations can be used to probe both their equilibrium and non-equilibrium dynamics.
While response and correlation functions are related in equilibrium by fluctuation-dissipation theorems, out of equilibrium they provide complementary information \change{on the dynamics}. 
In the past years, a method has been devised to map the quantum dynamics of an isolated extended system after a quench onto a static theory with boundaries in imaginary time; up to now, however, the focus was entirely on symmetrized correlation \change{functions}. Here we \change{provide a prescription which, in principle, allows one to retrieve}   
the whole set of relevant dynamical quantities characterizing the evolution, including linear response functions. We illustrate this 
construction with some relevant examples, showing \change{in the process} the emergence of 
light-cone effects similar to those observed in correlation functions.
\end{abstract}



\date{\today}

\maketitle


\begin{section}{Introduction}

Motivated by significant experimental advances in engineering and manipulating ultra cold atomic systems in optical lattices\cite{coldatoms}, the coherent dynamics of thermally isolated, spatially extended quantum systems has been recently the subject of intense theoretical and experimental 
investigation\cite{PSSV-11}. 
Perhaps the simplest protocol for studying dynamical properties is the so-called quantum quench: the system is prepared at time $t=0$ in the ground state $| \psi_0\rangle$  of a certain Hamiltonian $H_0$ and then one lets it evolve according to a globally different Hamiltonian $H$.

Time-dependent \emph{correlation functions} of suitable quantities after the quench reveal important features of the dynamics of the system: they not only form the natural basis of its theoretical description\cite{IR-00,K-06,CC-06a,CC-06b,CEF-12a,EEF-12,M-13} but are also experimentally 
accessible\cite{Cheneau2012,Jurcevic2014,Richerme2014}. 
In particular, the qualitative behavior of correlation functions can be understood in terms of entangled quasiparticles of $H$ which are generated upon quenching and which propagate across the system with a finite speed\cite{CC-06a,CC-06b}, giving rise to the "light-cone effects" observed in 
experiments\cite{Cheneau2012,Jurcevic2014,Richerme2014}. In addition, some features of correlation functions at large distances and late times are largely independent of the specific Hamiltonian
$H$, i.e., they are \emph{universal}, as long as $H$ is sufficiently close to quantum 
critical points\cite{CC-06a,CC-06b}. The consequences of $H$ being critical\cite{Sachdev} (with dynamic exponent $z=1$) are particularly important in one space dimension because the underlying conformal symmetry of the problem constrains the functional form of the time-dependent correlation functions. 
These conclusions were drawn 
by mapping the dynamics after the quench of the $d$-dimensional system  onto the static behavior of a $d+1$-dimensional near-critical system confined in a slab with proper boundary conditions, which can be analyzed via renormalization-group (RG) 
arguments\cite{CC-06a,CC-06b}. 
In doing so, one can take advantage \change{(see, e.g., Ref.~\onlinecite{CG-11})} of the available knowledge about the thermodynamics and structural properties of statistical systems confined by boundaries\cite{boundaries, boundaries1,boundaries-P} within slabs of finite thickness\cite{fss}.

Within this framework, imaginary time is nothing but a spatial coordinate and therefore the expectation values which can be calculated via this approach are typically symmetric upon exchanging times and they appear unable to encode the causality inherent to any dynamical response,
\change{as the latter has to vanish whenever the external perturbation is applied later than the time at which its effects on the system are measured.}
This would constitute a serious limitation, since the time-dependent (linear) response of the system to external perturbations provides a direct and natural way to probe its dynamics.
In thermal equilibrium, response functions are related to dynamic correlations by 
fluctuation-dissipation relations\cite{Kubo} and therefore they actually provide no independent insight into the dynamics.
This is no longer the case out of equilibrium where, instead, they can be used in order to probe the eventual thermalization or to define effective temperatures in classical\cite{C-11} and quantum\cite{FCG-11,FCG-12,KPSR-13,LOG-10} systems.
Accordingly, a thorough theoretical and experimental description of the dynamics out of equilibrium necessarily requires a joint study of correlations and response functions.
In spite of their relevance, however,  response functions after a quantum quench have received so far less attention than correlation functions, having been studied analytically or numerically only in few cases\cite{K-06,FCG-11,FCG-12,KPSR-13,EEF-12}.
In addition, as anticipated above, it would appear that they cannot be retrieved theoretically from the aforementioned mapping to imaginary time.

Here we show that this is not the case and, as a matter of fact, we provide a prescription for determining any of the dynamical (causal) functions emerging within the Keldysh formalism --- which was developed in the 60s\cite{K-65} 
for studying non-equilibrium processes in many-body systems\cite{K-book}.
While the mapping onto the slab relies
on a path-integral formalism involving an integration along imaginary times, the Keldysh formalism
is formulated entirely in real times, but it requires  the time integrals to be performed along a closed temporal contour, as discussed 
further below.
As applications, we discuss the response function of  generic systems in $d=1$ quenched at their critical points which follows from conformal field theory, exemplified with the Ising universality class. In $d>1$ we briefly consider the relevant case of the Gaussian model, i.e., the continuum limit of a $d$-dimensional lattice of (linearly-coupled) harmonic oscillators.

\end{section}


\begin{section}{Mapping to imaginary times}
\label{sec:mapping}

The expectation value  $\langle \ob (t) \rangle= \bra{\psi_0} \rme{iHt} \mathcal{O} \rme{-iHt} \ket{\psi_0}$ of an observable $\ob$ at time $t$, can 
be obtained from\cite{CC-06a,CC-06b}
\be
\langle \ob (t) \rangle_\epsilon \equiv Z^{-1} \bra{\psi_0} \rme{iHt-\epsilon H} \, \mathcal{O} \, \rme{-iHt -\epsilon H} \ket{\psi_0},
\label{eq:exp1}
\ee
as $\epsilon>0$ vanishes, where $Z = \bra{\psi_0} \rme{-2\epsilon H}  \ket{\psi_0}$ ensures normalization.
(The possible dependence of $\ob$ on the spatial coordinates is understood.)
The factors $\rme{-\epsilon H}$, introduced in order to ensure
the convergence of the path-integral representation of $\langle \ob (t) \rangle_\epsilon$,
can be regarded as evolution operators ``advancing'' the time by $-i\epsilon$ along the imaginary axis. 
$\langle \ob (t) \rangle_\epsilon $ can thus be expressed as
\be
\int\!\!\mathcal{D} \phi \, \bracket{\psi_0}{\phi(-i\epsilon)} \bracket{\phi(i\epsilon)}{\psi_0} \, \bra{\phi(t)}\mathcal{O} \ket{\phi(t)} \, \rme{i\int_{\gamma} \!\rmd t' \, L[\phi]},
\label{eq:exp2}
\ee
where the contour $\gamma$ is shown in Fig.~\ref{fig:conts}(a), $L$ is the Lagrangian corresponding to $H$, and $\phi$ is a complete set of fields (possibly including $\phi^\dag$) along $\gamma$.
By choosing different discretizations of the time-evolution operators in Eq.~\reff{eq:exp1},
$\gamma$ can be arbitrarily modified as long as it starts from $i\epsilon$ on the imaginary axis, proceeds downwards and rightwards till it reaches $t$ on the real axis and then continues downwards and leftwards to the final point $-i\epsilon$ on the imaginary axis; by introducing the identity 
$\rme{-iHt'}\rme{iHt'}$ with real $t'$ one can relax the constraint about the rightwards and leftwards orientation of the contour in the upper and lower complex half-plane, respectively; 
however, this consideration cannot be applied in general to
the downwards orientation because $\rme{H|\tau|}$ might be unbounded, hindering the construction of the path-integral representation.
%
%
%
\begin{figure}
\centering
\includegraphics[trim = 35mm 115mm 50mm 70mm, clip,scale=0.315]{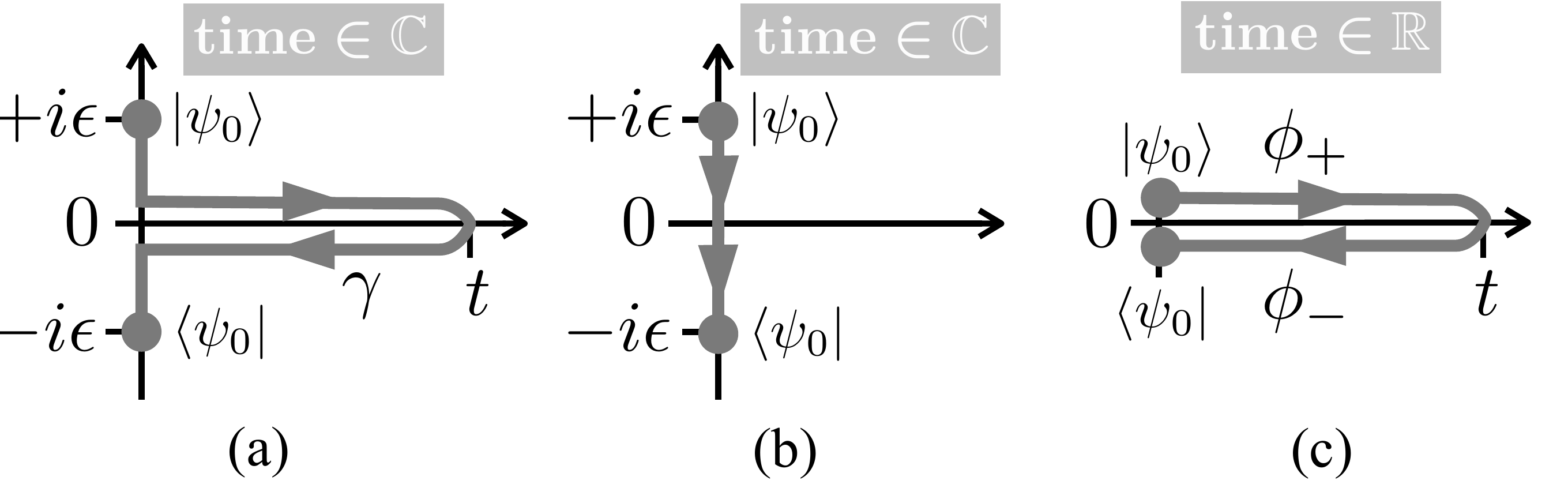} 
\caption{Contours of temporal integration of the Lagrangian $L[\phi]$ in the path-integral representation~\reff{eq:exp2} of expectation values. 
The initial and final points of the oriented contour $\gamma$ correspond to the ground state $|\psi_0\rangle$ of the initial Hamiltonian.
(a) Example of a contour for the evaluation of $\langle \ob (t) \rangle_\epsilon$ in Eq.~\reff{eq:exp1}: the introduction of the regulator $\epsilon>0$ allows one to consider a complex $t$ within the strip $|{\rm Im} \, t| \le \epsilon$. 
A vanishing ${\rm Re}\, t$ amounts at evaluating $\langle \ob (t) \rangle_\epsilon$ in Euclidean time, i.e., along the contour indicated in panel (b).
The contour in panel (c) involves only real times and is naturally introduced within the Keldysh formalism\cite{K-book} for representing $\langle \ob(t)\rangle$ after a quench from a pure initial state $|\psi_0\rangle$.
}
\label{fig:conts}
\end{figure}
%
These facts allow one to calculate $\langle \mathcal{O} (t) \rangle$ as discussed in Refs.~\onlinecite{CC-06a,CC-06b}: using the property of analyticity of Eq.~\reff{eq:exp1}, one can restrict to imaginary times, i.e., to the contour in Fig.~\ref{fig:conts}(b),
determine
\be
\langle \mathcal{O} (i\tau) \rangle_\epsilon \equiv Z^{-1} \bra{\psi_0} \rme{-(\epsilon + \tau) H} \, \mathcal{O} \, \rme{-(\epsilon -\tau ) H} \ket{\psi_0},
\label{eq:exp3}
\ee
for $|\tau| < \epsilon$ and 
eventually perform an analytic continuation $\tau \mapsto - it$. Equation~\reff{eq:exp3} can be represented as 
\be
\int\!\! \mathcal{D} \phi  \bracket{\psi_0}{\phi(-i\epsilon)} \bracket{\phi(i\epsilon)}{\psi_0} \bra{\phi(i\tau)}\mathcal{O} \ket{\phi(i\tau)} \rme{-\int_{-\epsilon}^{\epsilon} \!\rmd\tau'  L_E[\phi]},
\label{eq:exp4}
\ee
where $L_E$ is the Euclidean Lagrangian corresponding to $H$. As pointed out in Refs.~\onlinecite{CC-06a,CC-06b}, Eq.~\reff{eq:exp4} has the form of an equilibrium expectation value in a $(d+1)$-dimensional system within a slab with boundaries at $\tau=\pm\epsilon$ and the operator $\ob$ inserted at some $|\tau| < \epsilon$. The initial state $\ket{\psi_0}$ encodes the conditions at such boundaries.

This correspondence is particularly fruitful for 
critical systems in $d=1$ which, being exactly solvable in the slab due to their conformal symmetry, allow a straightforward continuation from imaginary to real times; conformally invariant boundary states $\ket{\psi_0^\ast}$, however, are in general not normalizable\cite{BCFT}. \change{On the other hand, this issue can be bypassed by approximating the latter with states $\ket{\psi_0}$ lying within their renormalization-group basin of attraction\cite{CC-06a,CC-06b}}.
As long as one is only interested in the leading scaling behavior at long times and large distances, \change{the relevant features of $\langle \ob (t) \rangle_\epsilon $ in Eq.~\reff{eq:exp1} are} effectively determined by the boundary state $\ket{\psi_0^\ast}$ which $\ket{\psi_0}$ flows to under RG transformations. 
If $\ket{\psi_0}$ is sufficiently close to $\ket{\psi_0^\ast}$, it gives rise to approximately equivalent boundary conditions, the main difference being that they are applied outside the slab at a distance $\tau_0$ from its actual boundaries. This distance is known as \emph{extrapolation length}\cite{boundaries, boundaries1}.
For $\epsilon \to 0$  the remaining effective slab has width $2 \tau_0$.
This analysis carries over to correlation functions involving different times and quantities; however, as $\gamma$ is always oriented downwards, the generalization of Eq.~\reff{eq:exp4} to this case renders the anti-time ordering $T^\ast$ along the imaginary axis of the corresponding observables, e.g.,
\begin{align}
	\av{T^\ast \lqq \ob(i\tau_1) \ob(i\tau_2) \rqq} &= \theta(\tau_2 - \tau_1) \av{ \ob(i\tau_1) \ob(i\tau_2)} + \nonumber \\
	&+  \theta(\tau_1 - \tau_2) \av{ \ob(i\tau_2) \ob(i\tau_1)},
	\label{eq:antiord}
\end{align}
where $\theta(t>0)=1$ and 0 otherwise.

\end{section}


\begin{section}{Keldysh approach}
\label{sec:Keldysh}

Alternatively to the mapping discussed in Sec.~\ref{sec:mapping}, 
$\langle \ob(t)\rangle$ can be represented directly as in Eq.~\reff{eq:exp2}  with the contour shown in Fig.~\ref{fig:conts}(c), which consists of a forward and a backward branch running along the real axis\cite{K-book}. In order to distinguish them, one introduces two fields $\phi_+(t')$ and $\phi_-(t')$ which stand for $\phi(t')$ along the former and the latter, respectively, with the condition $\phi_+(t) = \phi_-(t)$.
(In multi-time correlation functions, this condition is imposed at the largest time.)
In terms of $\phi_\pm$,
Eq.~\reff{eq:exp2} acquires the form of a path integral in which the action $\int_\gamma L[\phi] = \int_0^t \rmd t' (L[\phi_+(t')] -  L[\phi_-(t')])$ involves the usual forward integration in time.
Time-dependent correlation functions of $\phi_\pm$ --- which are time-ordered along the contour $\gamma$ --- can be obtained by standard methods, i.e., functional derivatives of the partition function, by introducing  fields $h_\pm$ conjugate to $\phi_\pm$ which generate a term $\int_0^t \rmd t' [h_+(t')\phi_+(t') -  h_-(t')\phi_-(t')]$ in the exponential of Eq.~\reff{eq:exp2}. 
Among the four two-time correlation functions of the fields $\phi_\pm$ which can be formed, we focus on 
\begin{subequations} \label{eq:gtrless}
\begin{align}
	&iG^>(t_1,t_2) = \langle T_K[\phi_-(t_1)\phi^\dagger_+(t_2)]\rangle \quad\mbox{and} \\
	&iG^<(t_1,t_2) = \langle T_K[\phi_+(t_1)\phi^\dagger_-(t_2)]\rangle,
	\label{eq:lgr}
\end{align}
\end{subequations}
where $T_K$ implements the ordering along the Keldysh contour $\gamma$. In these terms, the symmetrized correlation function 
$C_+(t_1,t_2) \equiv \bra{\psi_0} \{ \phi(t_1),\phi^\dagger(t_2)\}\ket{\psi_0}$ can be  expressed as $C_+(t_1,t_2) = i[G^<(t_1,t_2) + G^>(t_1,t_2)]$.

\end{section}


\begin{section}{The response function}
\label{sec:response}

The response function $R(t,s)$ describes the linear variation of a quantity at time $t$ (e.g., the expectation value 
$\langle \phi(t) \rangle$) due to a perturbation $h$ applied (somewhere else in space) at an earlier time
\be
R(t,s) = \left.\frac{\delta \langle \phi(t) \rangle}{\delta h(s)}\right|_{h\equiv 0}.
\ee
Clearly, $R(t,s)$ must vanish for $t<s$ because of causality, \change{i.e., no difference can be observed deriving from a perturbation which has not not yet been applied}.
In the case of a ferromagnet, for example, $\phi$ is the local magnetization and therefore $R(t,s)$ quantifies how much it changes at time $t$ by turning on a weak (local) magnetic field at time $s$.
Within the Keldysh approach discussed above in Sec.~\ref{sec:Keldysh}, 
one readily finds the Kubo relation\cite{Kubo,K-book}  (with $\phi(t)=\phi_\pm(t)$)
which holds both in and out of equilibrium
\be
R(t,s) = \theta(t-s) [G^<(t,s) - G^>(t,s)]
\label{eq:R}
\ee
\change{and which relates the response function to the correlation functions $G^\gtrless$ defined in Eq.~\reff{eq:gtrless}.} 
In what follows, \change{whenever times are indicated by $t$ and $s$,  we always assume $t>s$}.
The crucial point here is to understand how to extract $G^{\gtrless}$ from the two-point (static) correlation function in the slab.
As anticipated above, the variable 
(imaginary time) which has to be continued to imaginary values in order to recover the actual quantum evolution after the quench (see also Sec.~\ref{sec:mapping}) is akin to a spatial coordinate; thus, \change{no causal structure can be expected to emerge in this framework. In particular, due to the rotational symmetry along the axes running parallel to the axis of the imaginary time, the two-point functions that one can retrieve acquire some symmetry under the exchange of time coordinates: this is not compatible with the intrinsic temporal asymmetry of causal quantities such as $R$, which vanish identically only for a certain ordering of the time variables.}

We recall --- see Sec.~\ref{sec:mapping} --- that in order to have access to imaginary times, one has to introduce regularizing factors as in Eq.~\reff{eq:exp1}. This results in the Keldysh contour depicted in Fig.~\ref{fig:conts}(a), in which $\phi_+$ and $\phi_-$ are extended to the upper and the lower half planes, respectively.
As discussed in Sec.~\ref{sec:mapping}, the correlations calculated via the Euclidean path integral in Eq.~\reff{eq:exp4} and the contour in Fig.~\ref{fig:conts}(b) are effectively anti-time-ordered along the imaginary axis. Furthermore, such an ordering cannot be altered by deforming the contour for the purpose of the analytic continuation; accordingly, accounting for the order of the fields in $iG_E (\tau_1, \tau_2) \equiv \langle T^\ast [\phi(i\tau_1) \phi^\dag(i\tau_2)] \rangle_\epsilon$ \change{(see Eqs.~\reff{eq:antiord} and \reff{eq:lgr})}, \change{one sees that the analytic continuation of $iG_E$ for $\tau_2 > \tau_1$ renders $G^>$ whereas the one for $\tau_2 < \tau_1$ yields $G^<$.} 
This results in the prescription 
\be
G^{\gtrless} (t_1, t_2) =  G_E (-it_1 \mp 0^+   , -i t_2),
\label{eq:GE}
\ee
which allows the calculation of the linear response function $R$ according to Eq.~\reff{eq:R}.
In particular, from it we can read that $R$ 
vanishes identically 
if the analytic continuation of $G_E$ to real times is single-valued.

\end{section}


\begin{section}{Applications}

\change{In order to illustrate and exemplify the general discussion presented in the previous sections, we consider below the non-equilibrium dynamics of generic critical quantum systems in one spatial dimension and of the Gaussian model in higher-dimensions.}

\begin{subsection}{Critical systems in $d=1$}

\label{sec:applic-1dcrit}

\change{According to the discussion of Sec.~\ref{sec:mapping}, the dynamics of a quantum systems in $d=1$ quenched at a critical point with dynamical exponent $z=1$ can be studied in terms of a $d=2$ conformal theory bounded within a strip, with suitable boundary conditions.\cite{CC-06a,CC-06b}} The Euclidean correlation function $iG_E$ of the order parameter $\phi$ is 
\be
\langle \phi(r,\tau_1) \phi (0,\tau_2) \rangle = \xi^x F(\eta),
\label{eq:iGE-gen}
\ee
where $x$ is the scaling dimension of the (primary) field $\phi$ (assumed here to be scalar, for simplicity),
\be
\label{eq:2p}
\xi = \left(\frac{\pi}{2\tau_0}\right)^2 \frac{\cosh r +\cos(\tau_1 + \tau_2)}{4\cos \tau_1 \cos \tau_2 [\cosh r - \cos(\tau_1 - \tau_2)]},
\ee 
$\tau_0$ is the extrapolation length, $r>0$ the spatial distance between the two points and $\tau_{1,2}$ the corresponding imaginary times within the strip  $|\tau_{1,2}| < \pi/2$. Distances and times are given here in units of $2\tau_0/\pi$ \change{(e.g., $r \to \pi r /2\tau_0$)}. 
The function $F$ in Eq.~\reff{eq:iGE-gen} depends on
\be 
	\eta = \frac{2\cos \tau_1 \cos \tau_2}{\cosh r + \cos(\tau_1+ \tau_2)} \le 1,
	\label{eq:harmR}
\ee
on the boundary states, and on the \change{specific critical system under consideration} but \change{displays (after a suitable normalization of the field)} the generic properties $F(1)=1$ and $F(\eta\ll 1)\propto \eta^{x_b}$, where $x_b$ is the scaling dimension of the most relevant boundary operator \change{appearing in the short-distance expansion of $\phi$} \cite{C-84, CC-06b, boundaries1}. 
For the Ising \change{universality class}, e.g., $x=1/8$ and\cite{C-84,CC-06a,CC-06b}
\be
F(\eta) = \sqrt{\frac{1+\sqrt{\eta}}{2}} \pm \sqrt{\frac{1-\sqrt{\eta}}{2}},
\label{eq:F}
\ee
for fixed $(+)$ and free $(-)$ boundary conditions at the edges of the slab. 

Due to the apparent symmetry $\tau_1 \leftrightarrow \tau_2$ of Eqs.~\reff{eq:2p} and \reff{eq:harmR}, the prescription  $\tau_{1,2} \mapsto - i t_{1,2}$ of Refs.~\onlinecite{CC-06a,CC-06b} can only be 
used for determining the asymptotic properties of the two-time (symmetrized) correlation function 
\change{$C_+(t_1,t_2) = u^x F(n)$} after the quench: \change{in fact, the expressions $u \equiv \xi(\tau_{1,2} = - i t_{1,2})$ and $n\equiv \eta(\tau_{1,2} = - i t_{1,2})$ \change{obtained from Eqs.~\reff{eq:2p} and \reff{eq:harmR}, respectively,} are symmetric under exchange of the times $t_{1,2}$:}
\begin{subequations}
\label{eq:un}
\begin{align}
	u &= \left(\frac{\pi}{2\tau_0}\right)^2 \frac{\cosh r +\cosh(t_1 + t_2)}{4\cosh t_1 \cosh t_2 [\cosh r - \cosh(t_1 - t_2)]}, \label{eq:u}\\
	n &= \frac{\cosh(t_1 - t_2) +  \cosh(t_1 + t_2)}{\cosh r + \cosh(t_1+ t_2)}.  \label{eq:n}
\end{align}
\end{subequations}
As a function of $r$ for fixed times $t_{1,2}$, $C_+$ is constant and $\propto \rme{-x |t_1-t_2|}$ for $r\ll |t_1-t_2|$, decays exponentially $\propto \rme{- x r}$ for $|t_1-t_2|\ll r \ll t_1+t_2$ while it is proportional to $\rme{- x_b r - x(t_1+t_2)}$ for $r\gg t_1+t_2$.

In order to determine $G^\gtrless$, \change{one must instead carefully perform the analytic continuation from within a specific domain:} according to Eq.~\reff{eq:GE} one has to continue $G_E$ towards imaginary values of its arguments (i.e., to real times) $\tau_1 \mapsto - i t_1 \pm 0^+$ and $\tau_2 \mapsto -i t_2$. 
In doing this,
$\cos(\tau_1- \tau_2)$ in $\xi$ [see Eq.~\reff{eq:2p}] turns into $\cosh(t_1-t_2\pm i0^+)$ in its real-time counterpart $u$ [see Eq.~\reff{eq:u}], 
such that $u^x$ (with non-integer $x$) displays a branch point and acquires an imaginary part for $|t_1-t_2| > r$. 
The continuation 
$n = \eta(\tau_{1,2} = -it_{1,2})$ of $\eta$ [see Eqs.~\reff{eq:n} and \reff{eq:harmR}], 
instead, lies within the domain of analyticity of $F$ for $|t_1-t_2|<r$ [see Eq.~\reff{eq:n}] and therefore the continuation $\widehat F(n)$ of $F(\eta)$ does not develop an imaginary part within this range.
This implies via Eq.~\reff{eq:R} that the response function $R(t,s)$ vanishes identically for $t-s < r$, whereas it does not for $t-s>r$: Accordingly, $R$ displays a light-cone effect similar to the one numerically observed in a previous study.\cite{FCG-12} 
This can be reinterpreted in terms of \change{a quasi-particle picture analogous to the one originally introduced for correlation functions}\cite{CC-06a,CC-06b}: in the present case, the excitations produced locally by the perturbation applied at time $s$ and propagating with unit velocity 
need at least a time $r$ to cover the distance $r$ at which the effect of the perturbation is measured. 

In the scaling limit $t_{1,2}, r\gg 1$ (with $|t_1-t_2|>r$), it turns out that $n \to 1$ and $u \propto - (\rme{|t_1-t_2|}-\rme{r})^{-1} $; therefore, $R$ decays as 
\be
R(t>s+r,s\gg 1) \propto \rme{-x(t-s)} \quad \mbox{for} \quad  t-s \gg r,
\label{eq:Rdecay}
\ee
analogously to $C_+$,
while it displays an algebraic singularity $\propto (t-s-r)^{-x}$ for $t \to s + r$, independently of the initial condition encoded in $F$.
For the Ising \change{universality class} one finds \change{from the analytic continuation of Eq.~\reff{eq:F} (see the Appendix~\ref{app:A} for details) that the response function $R$ for $t-s > r$ is given by} 
\be
\begin{split}
&R(t,s) = 
 \sqrt{2} \abs{u}^{1/8} \\[2mm]
&\quad\times \left[ \sina{\frac{\pi}{8}} \sqrt{n^{1/2} + 1}  \mp \cosa{\frac{\pi}{8}} \sqrt{n^{1/2} - 1}  \right],
\end{split}
\label{eq:RI}
\ee
with $u$ and $n$ as in Eq.~\reff{eq:un}. \change{The plot of $R(t,s)$ in provided in Fig.~\ref{fig:RR} as a function of $t-s$ for various values of $s$, $r$, and fixed and free boundary conditions.}
%
%
\begin{figure}
\centering
\includegraphics[trim = 28mm 23mm 125mm 83.9mm, clip,scale=0.31]{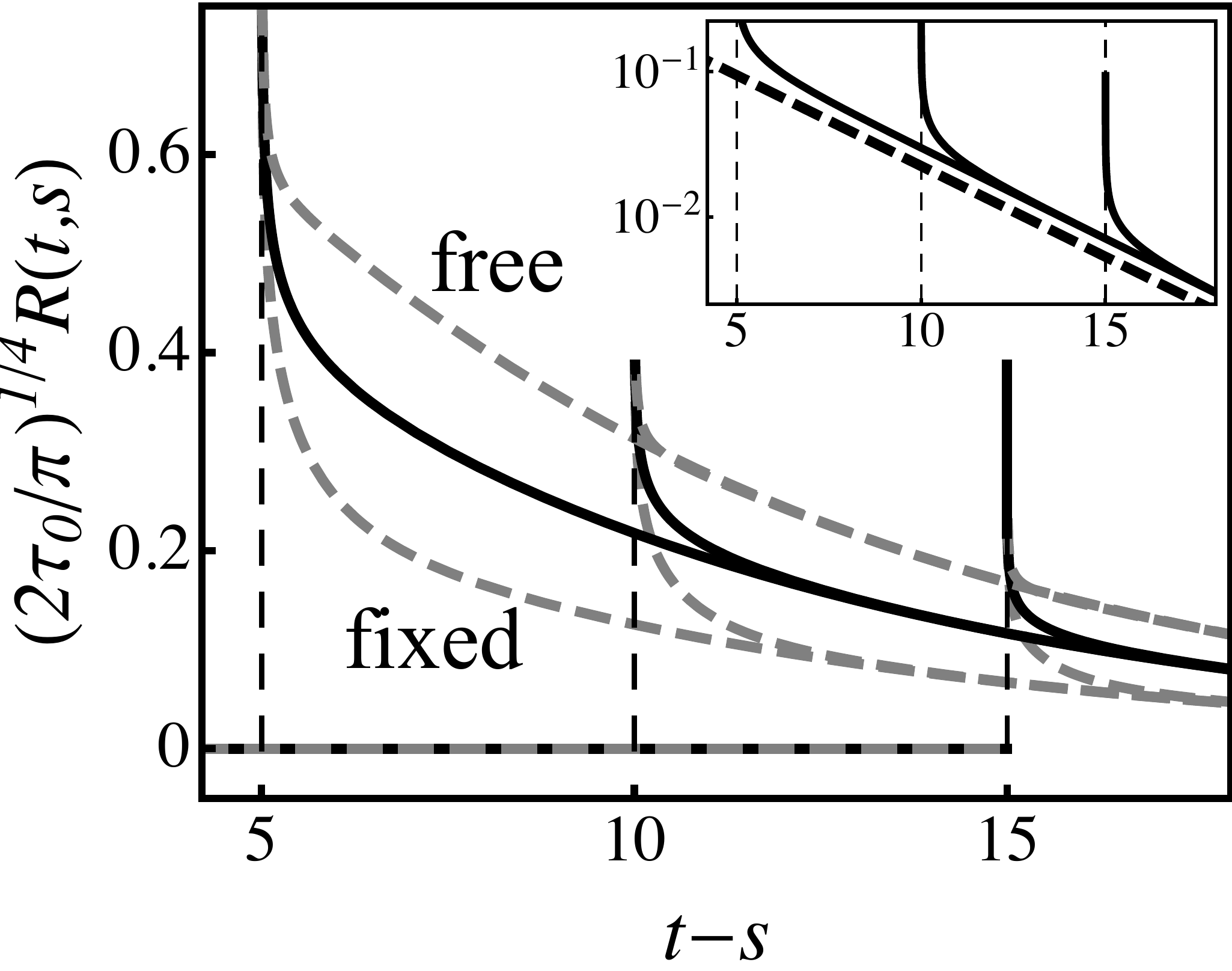} 
\caption{Response function $R(t,s)$ of the order parameter of the $d=1$ quantum Ising universality class after a quench to the critical point, as a function of $t-s$, for $s=1$ (dashed) and $s=5$ (solid), with $r=5$, 10, and 15. 
With fixed $r$, $R$ vanishes for $t-s<r$ and displays an algebraic divergence for $t-s \to r^+$. The upper and lower sets of dashed curves correspond to free and fixed boundary conditions, respectively. As $s$ increases beyond $\simeq 5$, $R$ becomes effectively independent of the boundary conditions (solid line). The inset shows the corresponding $R$ in a logarithmic scale and highlights its long-time exponential decay $\propto \rme{-(t-s)/8}$ (thick dashed line). \change{In these plots, times are given in units of $2\tau_0/\pi$, where $\tau_0$ is the value of the extrapolation length which characterizes the initial state of the quench (see the main text).}}
\label{fig:RR}
\end{figure}

\begin{figure}[ht]
\centering
\includegraphics[clip, scale = 0.56]{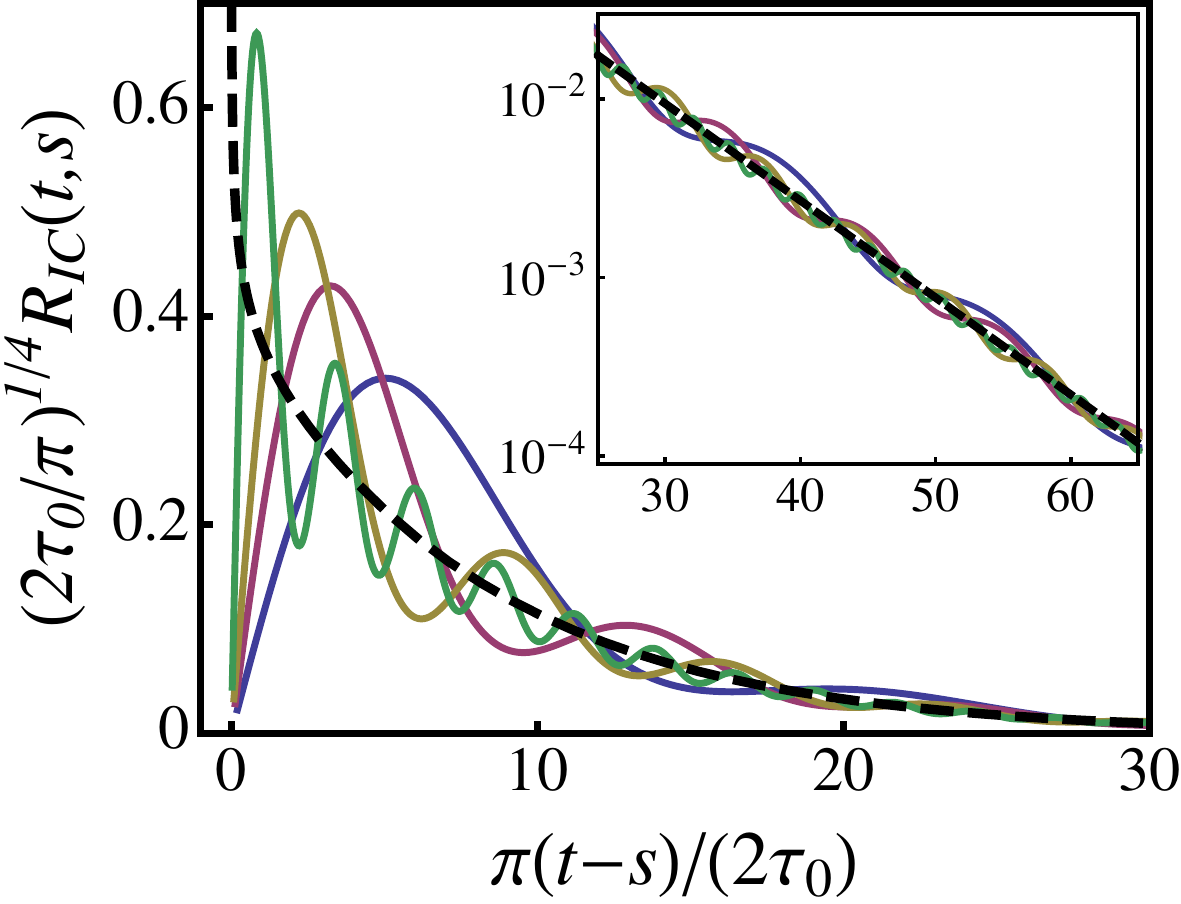}
\caption{(Color online) Response function $R_{IC} (t,s)$ of the order parameter of the quantum Ising chain after a quench of the transverse field strength $\Gamma$ which ends at the critical point $\Gamma=\Gamma_c=1$. 
The response is measured at the same point at which the perturbation is applied (corresponding to $r=0$) and $t$ and $s$ are chosen large enough for $R_{IC}$ to become stationary. The various solid lines represent the numerical data of Ref.~\onlinecite{FCG-12} for different choices of the initial state, which have been rescaled with the values of $\tau_0$ determined from the exponential slope $\propto \rme{-\pi (t-s) / (16 \tau_0)}$ observed for $t-s \gg r$ [see Eq.~\reff{eq:Rdecay} in which the time units have been reinstated]. 
\change{With an increasing slope in the origin, the solid lines refer to $\Gamma_0 = 0$, 0.3, 0.5, and 0.8.}
The dashed line, instead, corresponds to the rescaled theoretical prediction in Eq.~\reff{eq:RI}. The inset shows the same curves as in the main plot but in a logarithmic scale and for a wider range of times.}
\label{fig:RIC}
\end{figure}
%
\change{The qualitative features of $R$ in Eq.~\reff{eq:RI} are consistent with those emerging from a numerical calculation\cite{FCG-12} of the response function $R_{IC}$ of the quantum Ising chain in a transverse field of strength $\Gamma$, long after a quench  to the critical point $\Gamma=\Gamma_c\equiv 1$, starting from the  ground state corresponding to $\Gamma=\Gamma_0<1$.  
In Fig.~\ref{fig:RIC} we compare $R_{IC}(t,s)$ in the stationary regime $s \gg 1$ (see Figs.~12 and 13 of Ref.~\onlinecite{FCG-12}) with the analytical prediction \reff{eq:RI}; both quantities refer to the case $r=0$ in which the effect of the perturbation on the order parameter (to which it couples linearly) is measured at the same spatial point where the perturbation was applied.} 
\change{In order to compare the numerical data of Ref.~\onlinecite{FCG-12} with Eq.~\reff{eq:RI} [see also Eq.~\reff{eq:un}] it is first necessary to determine the value $\tau_0(\Gamma_0)$  of the extrapolation length corresponding to a certain initial state of the quench, which is characterized by the transverse magnetic field $\Gamma_0$ in the numerical calculation. This can be done by identifying the rate of the exponential decay of $R_{IC}$ as a function of $t-s$ (which depends on $\Gamma_0$) with  $16\tau_0(\Gamma_0)/\pi$, as predicted by Eq.~\reff{eq:Rdecay} with $x=1/8$, after reinstating the time units.}
Once $\tau_0(\Gamma_0)$ is determined, a data collapse of the curves corresponding to different values of $\Gamma_0$ should be observed by plotting $\tau_0^{1/4}(\Gamma_0)R_{IC}$ as a function of $(t-s)/\tau_0(\Gamma_0)$, as predicted by Eq.~\reff{eq:RI} in the stationary regime, in conjunction with Eq.~\reff{eq:u}. 
Up to oscillations due to lattice effects \change{(studied in some detail in Refs.~\onlinecite{FCG-11,FCG-12})}, 
Fig.~\ref{fig:RIC} shows that indeed this collapse occurs (curves corresponding to various values of $\Gamma_0$ are indicated by solid lines of different colors): in fact, \change{these rescaled curves share not only the rate of the long-time exponential decay, but also the overall amplitude}, as highlighted in the inset. The dashed line indicates the master curve predicted on the basis of Eq.~\reff{eq:RI}, up to an overall multiplicative factor which has been determined in order to match the (common) asymptotic behavior of the numerical curves. Note, in fact, that in comparing the correlation or response functions of a continuum theory with those of a lattice model, one has always to allow for an arbitrary overall amplitude due to the possibly different normalizations of the fields. 
The same scaling with $\tau_0$ and a qualitative agreement with the theoretical predictions such as that illustrated here for $R_{IC}$ with $r=0$  is found also for the response function with $r\neq 0$ in the stationary state.
In particular, the data collapse of the rescaled curves at long times (as in Fig.~\ref{fig:RIC}) can be obtained also in this case by using the same values of $\tau_0(\Gamma_0)$ as determined here. This confirms the fact that $\tau_0$ is the only relevant scale for determining the leading critical behavior which enters the response function.
\change{The time-scale of the exponential decay --- proportional to $\tau_0(\Gamma_0)$ --- determined from the numerical data as explained 
above is accurately described --- at least for $\Gamma_0=0$, 0.3, 0.5, and 0.8 --- by Eq.~(86) in Ref.~\onlinecite{FCG-12} (see also its derivation in Refs.~\onlinecite{CEF-11,CEF-12a}), i.e., with the current notation, $\tau_0= (\pi/8)^2\sqrt{\Upsilon-1}/\arctan(\sqrt{\Upsilon-1})$, where $\Upsilon = [(1+\Gamma_0)/(1-\Gamma_0)]^2$.
The increase $\propto (1-\Gamma_0)^{-1}$ of this $\tau_0(\Gamma_0)$ for $\Gamma_0\to 1$ is compatible with the heuristic idea that upon decreasing the amplitude of the quench, $|\psi_0\rangle$ becomes increasingly different from $|\psi^*_0\rangle$, as signaled by the increasing extrapolation length.
As a matter of fact, the latter turns out to reproduce qualitatively the behaviour of the correlation length of the initial state\cite{CC-06b, boundaries1}.} 

In passing we mention that  in the stationary regime $t,s\gg r$, $C_+$ and $R$ for the Ising universality class \change{determined according to the prescription discussed in Sec.~\ref{sec:response} and on the basis of Eqs.~\reff{eq:iGE-gen}, \reff{eq:2p}, \reff{eq:harmR}, and \reff{eq:F} (see also Appendix~\ref{app:A})} 
turn out to satisfy the fluctuation-dissipation theorem ${\rm Im\,} \widetilde R(\omega) = \tanh(\pi\omega) \widetilde C_+(\omega)$ \change{--- where $ \widetilde R$ and $\widetilde C_+$ indicate the Fourier transforms of $R(t_2 - t_1)$ and $C_+(t_2 - t_1)$, respectively ---} 
confirming the apparent thermalization of conformal models\cite{CC-06a,CC-06b} which is however due to the peculiar features of the initial state.\cite{CEF-12b}

\end{subsection}

\begin{subsection}{Gaussian model}
\label{sec:applic-Gauss}

As a simple but relevant example in space dimensionality $d>1$, 
consider the Gaussian model on the continuum with $L_E = \int \rmd^d x [ (\partial_\tau \phi)^2 + (\vec{\nabla} \phi)^2  + m^2  \phi^2 ]/2$ and "mass" $m$, which forms the basis of many approximations to the behavior of actual physical systems.
For isotropic initial conditions, its two-point function in the slab is\cite{boundaries1}
\be
\begin{split}
&i G_E (\vec{k}; \tau_1,\tau_2) = \frac{\rme{-\omega_k | \tau_1-\tau_2 | }}{2\omega_k} + A_k \rme{-\omega_k(\tau_1+\tau_2)} \\
&\quad\quad\quad\quad+ 2B_k \cosh(\omega_k(\tau_1-\tau_2)) + C_k \rme{\omega_k(\tau_1+\tau_2)},
\end{split}
\label{eq:iGE-gaux}
\ee
\change{as a function of the $d$-dimensional wave-vector $\vec{k}$ parallel to the slab boundaries (along which translational invariance holds) and of the distances $\tau_{1,2}$ of the two points from the center of the slab in the transverse direction.} In this expression
$k = |\vec{k}|$ and $\omega_k =\sqrt{m^2 + k^2}$ is the dispersion relation, while $A_k$, $B_k$ and $C_k$ depend on the initial condition and vanish when the latter is the ground state of the model; 
their specific values are not relevant for the present discussion because $R$ turns out to be independent of them.
In fact, the only non-analytic term in $i G_E$ --- which provides different results upon continuation to real times --- is the first one.
Due to the non-analiticity of $|\tau_1-\tau_2|$ for $\tau_1=\tau_2$, Eq.~\reff{eq:GE} corresponds to continuing the two different functions $\rme{\pm \omega_k (\tau_1 - \tau_2)}$ on the r.h.s. of Eq.~\reff{eq:iGE-gaux}; this yields 
\be
R(t, s) =   \frac{\sin(\omega_k (t-s))}{\omega_k},
\label{eq:R-Gaux}
\ee
in agreement with the result calculated directly within the Keldysh approach.\cite{K-book} %
In the coordinate representation with position $\vec{r}$ obtained by taking the Fourier transform of Eq.~\reff{eq:R-Gaux} with respect to $\vec{k}$, 
a branch point such as the one discussed for $d=1$ emerges clearly for
$m=0$ (though present also for $m\neq 0$):  in fact
\be
R(t, s) = \frac{[(t - s)^2 - r^2 ]^{-(d-1)/2}}{2\pi^{(d-1)/2} \Gamma((3-d)/2)} \theta(t-s-r).
\label{eq:RG}
\ee 
Also in this case, the relevant dynamical quantities in real time are encoded in the analytic structure of 
the correlation $i G_E$ within the slab; in particular, due to the presence of branches and singularities, the analytic continuation of $i G_E$ to the real axis provides different results when performed from different domains. As the two examples above highlight, each domain is characterized by a certain ordering of the imaginary parts of the time coordinates.
In passing, note that $R$ in Eq.~\reff{eq:RG} apparently vanishes for odd $d\ge 3$,
while the branching points simultaneously become simple or multiple poles.
In this case, the limit implied by Eq.~\reff{eq:GE} has to be understood in the sense of distributions and correspondingly $R$ displays an extremely sharp light-cone effect, being entirely concentrated on the horizon (see, e.g., the case $d=3$ detailed in Appendix B). 
\change{Note that this highlights a slightly peculiar property, as in odd dimension the light-cone effect becomes pronounced to the point that the response is non-vanishing only on the horizon itself; the introduction of interactions, i.e., of non-linear terms in the fields, however, is generically expected to smooth this behaviour and broaden the support of the response around the light cone.} 
\change{In fact, this feature is absent in the one-dimensional critical models discussed in Sec.~\ref{sec:applic-1dcrit}, which indeed describe (strongly) interacting systems.}

\end{subsection}

\end{section}


\begin{section}{Conclusions}

Out of equilibrium, correlation functions do not provide a complete description of the dynamics, making the 
knowledge of response functions necessary. The Euclidean approach devised for studying quantum quenches had been employed for determining the former\cite{CC-06a,CC-06b}; by mapping the dynamical problem onto a static one within a slab, the additional symmetry which emerges upon exchanging the (imaginary) time coordinates seemingly precludes the possibility of determining quantities --- such as response functions --- with a causal structure. 
Here we have shown that, on the contrary, these quantities can be retrieved by accounting for the analytic structure of static correlation functions within the slab. In particular, we demonstrated how to infer causal functions in quantum systems after a quench, working out 
the general features of the response function of one-dimensional quantum systems quenched to conformal critical points --- 
exemplified by the Ising chain in a transverse field (see Fig.~\ref{fig:RR}) --- and of the Gaussian model in higher spatial dimensionality.
Rather generically, the corresponding response functions display sharp light-cone effects which are analogous to those theoretically 
predicted\cite{CC-06a,CC-06b} and experimentally observed\cite{Cheneau2012,Jurcevic2014} for correlation functions: this describes the fact that information travels across the system at a finite speed ($v = 1$ in our units) and therefore one must wait for a local perturbation to propagate up to the considered point before being able to observe any response there.
The insight provided here on response functions is a step forward in our understanding of the generic and universal features of the dynamics after quantum quenches which are within the experimental reach.

\end{section}

\acknowledgments

We wish to thank P. Calabrese and A. Mitra for useful discussions.



\appendix

\begin{section}{Conformal Ising model}
\label{app:A}

\change{Here we provide some details on how to derive the expression in Eq.~\reff{eq:RI}
for the response function $R(t,s)$ of the Ising universality class, starting from the knowledge of the (Euclidean) correlation function
\be
iG_E = \xi^{1/8} F(\eta)
\ee
in the slab where $\xi$ and $\eta$ are as in Eqs.~\reff{eq:2p} and \reff{eq:harmR}, while 
$F(\eta)$ as in Eq.~\reff{eq:F}.} 
First of all, note that by continuing the variables $\tau_{k}$ ($k=1, 2$) to complex values $\tau_{k} - it_{k}$ (i.e., by adding an imaginary part) the various cosines appearing in Eq.~\reff{eq:2p} of $\xi$ transform according to
\be
	\cos \tau_k \to  \cos\tau_k \cosh t_k + i  \sin \tau_k	\sinh t_k.   \label{eq:cos} 
\ee
Since times and distances are measured in units of $2\tau_0/\pi$ (such that, e.g., actual distances $\widehat{r}$ are given by $\widehat{r} = 2\tau_0 r/\pi$), the boundaries of the slab are located at $\tau_k = \pm \pi/2$; in turn, this implies that within the allowed domain $|\tau_k| < \pi/2$ none of the factors inside $\xi$ vanishes, with the exception of
\be
	\cosh r - \cos(\tau_1 - \tau_2) \cosh(t_1 - t_2) + i  \sin(\tau_1 - \tau_2) \sinh(t_1 - t_2),
\ee
which instead does so for $\tau_1 = \tau_2 = 0$ and $\abs{t_1 - t_2} = r$.
	\begin{figure}[ht]
	\centering
	\vspace{1mm}
	\includegraphics[clip, width=0.9\columnwidth]{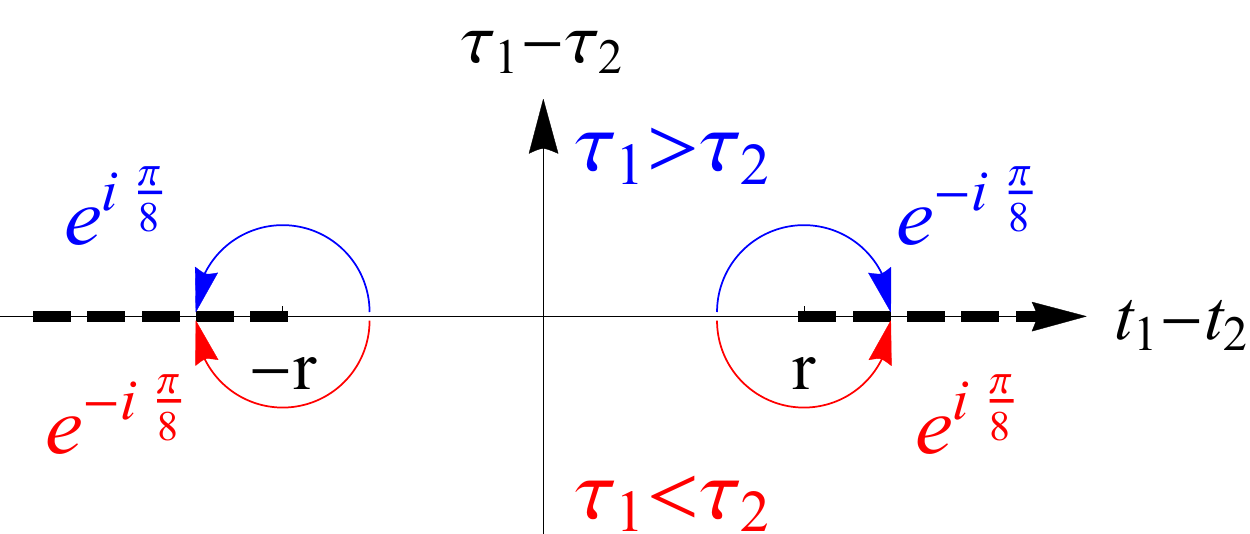}
	\caption{(Color online) Schematic representation of the analytic structure of the function $\xi^{1/8}$ [Eq.~\reff{eq:2p}] in the complex plane $z_1-z_2 \equiv (t_1 - t_2) + i(\tau_1 - \tau_2)$. The branch points are positioned at $\abs{t_1 - t_2} = r$; our (conventional) choice for the branch cuts is indicated by the thick, dashed lines superimposed to the real axis. The lower-half plane (red) represents the sector associated with $G^>$, whereas the upper half-plane (blue) with $G^<$. One can see that for $\abs{t_1 - t_2} > r$ the phase of $u^{1/8}$ depends on the choice of the imaginary ordering $\tau_1 \gtrless \tau_2$.}
	\label{fig:AACC}
	\end{figure}
\change{This means, e.g., that fixing $\tau_2 = 0$ and $t_2$ at some value on the real axis there are two points, lying on that same axis, at which $\xi$ becomes singular
as a function of $z_1\equiv t_1 + i \tau_1$.} 
Due to the fractional power $1/8$ of $\xi$, these become branching points for $i G_E$, as qualitatively sketched in Fig.~\ref{fig:AACC}: in fact, the analytic continuation of \change{$\xi^{1/8}$ --- and therefore of the 
two-point functions --- acquires different phases when $z_1-z_2$ approaches the real axis from either above or below [with $|\xi^{1/8}| = |u|^{1/8}$, see Eq.~\reff{eq:u}],}
therefore identifying a first difference between $G^<$ and $G^>$ retrieved as indicated by Eq.~\reff{eq:GE}.
Secondly, $\eta$ in Eq.~\reff{eq:harmR} can be safely continued 
as a function of $z_1$ and $z_2$ onto
the real axis, yielding $n$ in Eq.~\reff{eq:n}. 
Note that $n \le 1$ only if $\abs{t_1-t_2} \leq r$. Accordingly, in this case, the analytic continuation of the function $F$ in Eq.~\reff{eq:F} (which is originally defined only for $n \le 1$) is directly provided by $F(n)$ itself, whereas it becomes double-valued for $n>1$, i.e., $\abs{t_1-t_2}>r$; in particular, one finds that
\be
	\widehat{F}(n) = \sqrt{\frac{ 1 + \sqrt{n} }{2}}  \mp i \,\sigma  \sqrt{\frac{ \sqrt{n} - 1 }{2}}
\ee
with $\sigma \equiv {\rm sign} (t_1-t_2)$ if the real axis is approached from above, e.g., when determining $G^<$, while one finds $\widehat{F}^\ast$ if this approach occurs from below, as in the case of $G^<$. Hence we conclude that, for $\abs{t_1-t_2} > r$, 
\be
	iG^< (t_1,t_2) = \abs{u}^{1/8} \rme{i\pi\sigma/8} \widehat{F} (n)
\label{eq:gl1}
\ee
with $u$ given in Eq.~\reff{eq:u}. Note that for $t_1 + t_2 \gg r \gg 1$ and $\abs{t_1 - t_2} \gg r \gg 1$ the expression above behaves approximately as $\abs{u} \propto \rme{- \abs{t_1 - t_2}}$, whereas $n \simeq 1$; thus, the asymptotic behaviour of $G^<$ in this regime is dominated by the exponential $\rme{-\abs{t_1 - t_2}/8}$, which, reinstating the becomes $\rme{-\pi \abs{t_1 - t_2}/(16 \tau_0)}$. 
In view of this asymptotic behavior, one can estimate the extrapolation length $\tau_0$ by numerically or experimentally studying the large-time decay of these quantities, as exemplified above when discussing Fig.~\ref{fig:RIC}.
As $i G_E$ is real and invariant under the "space" reversal $\tau_k \mapsto - \tau_k$ (due to the symmetry of the slab and of the boundary conditions) one concludes that its analytic continuation satisfies $iG_E(-it_1 - 0^+,-it_2)  = [iG_E(-it_1 + 0^+,-it_2)]^\ast$, i.e., $i G^>(t_1,t_2) = [i G^<(t_1,t_2)]^\ast$. \change{Accordingly, for $t>s$ one has $R(t,s) = 2\, {\rm Im}\, i G^<(t,s)$, and by using the explicit expression of $iG^<$ in Eq.~\reff{eq:gl1}
one readily finds Eq.~\reff{eq:RI}.}

\end{section}

\begin{section}{Gaussian model}
\label{app:B}

\change{Here we focus on the case of the Gaussian model discussed in Sec.~\ref{sec:applic-Gauss}. As we are interested in determining first the Euclidean two-point function $iG_E$ in a slab, it is convenient to express it as a function of the $d$-dimensional wave vector $\vec{k}$ parallel to the confining surfaces} \change{(assumed to impose translationally invariant boundary conditions)} \change{and  of the distances $\tau_{1,2}$ of the two points from the middle plane of the slab. In this mixed representation, $iG_E$} \change{(away from the boundaries)} satisfies the equations $( \partial_{\tau_1}^2 - \omega_k^2) iG_E(\vec{k},\tau_1,\tau_2) = ( \partial_{\tau_2}^2 - \omega_k^2) iG_E(\vec{k},\tau_1,\tau_2) = -\delta (\tau_1 - \tau_2) $ (where $\omega_k = \sqrt{\vec{k}^2 + m^2}$) which are solved by Eq.~\reff{eq:iGE-gaux}
where the coefficients $A_{k}$, $B_{k}$, and $C_{k}$ are determined by imposing the proper boundary conditions on $iG_E(\vec{k},\tau_1,\tau_2)$. 
Even without knowing them, however, it is clear that the only term which might develop a point of non-analyticity 
is the first one on the r.h.s. of Eq.~\reff{eq:iGE-gaux} which in the following is referred to as $iG_{0E}(\tau\equiv \tau_1-\tau_2, \vec{k})$ and which in fact coincides with the two-point function 
of the Gaussian model in its ground state. 
The easiest way to derive Eqs.~\reff{eq:R-Gaux} and \reff{eq:RG} from Eq.~\reff{eq:iGE-gaux} is to consider the Fourier transform $iG_{0E} (\omega, \vec{k})$ of $iG_{0E}(\tau, \vec{k})$ in time:
\be
	iG_{0E} (\omega, \vec{k}) = \frac{1}{\omega^2 + \omega_k^2} = \frac{1}{\omega^2 + k^2 + m^2}
\ee 
which makes it apparent that the imaginary time $\tau$ is akin to every other spatial coordinate. By introducing the generalised vectors $\vec{\kappa} = (\omega, \vec{k})$ and $\vec{R} = (\tau,\vec{r})$ one finds (see, e.g., 10.32.10 in Ref.~\onlinecite{NIST})
\be
\begin{split}
	iG_{0E} (\tau, \vec{r}) &= \int \!\dd{d+1}{\kappa} \frac{\rme{i\vec{\kappa} \cdot \vec{R}}}{\kappa^2 + m^2}\\ 
	&=\frac{1}{(2\pi)^{(d+1)/2}} \left( \frac{m}{R}\right)^{(d-1)/2} K_{(d-1)/2}(mR)
	\label{eq:G0supp}
\end{split}
\ee
for $d>1$, where $K_\nu (z)$ is the modified Bessel function of the second kind. This function has generically a cut for ${\rm Arg\,} z = \pm \pi$ and is such that\cite{NIST} $K_\nu(z\to 0) = 2^{\nu-1}\Gamma(\nu) z^{-\nu}$, which renders Eq.~\reff{eq:RG} in the critical limit $m\to 0$.
Since $R^2 = r^2 + \tau^2$, it is not difficult to see that Eq.~\reff{eq:G0supp} has two branching points of algebraic order $(d-1)/2$ at $\tau = \tau_1 - \tau_2 = \mp ir$. Thus, its analytic structure is almost completely analogous to the one encountered when discussing the Ising case in Sec.~\ref{sec:applic-1dcrit} and in Appendix~\ref{app:A}, illustrated in Fig.~\ref{fig:AACC}, the only difference being in the acquired phase factors, which in the present case are $\rme{\pm i\pi (d-1)/2}$. 
Indeed, one finds that $iG^< = iG^>$ for $\abs{t_1 - t_2} < r$; for the massless case $m = 0$ and  $\abs{t_1 - t_2} > r$ one has, instead, 
\be
\begin{split}
&	iG^{\gtrless} (\vec{r},t_1,t_2) = \\
&	\quad \quad \quad \mal{K}_d \abs{r^2 - (t_1 - t_2)^2}^{-(d-1)/2} \rme{\mp i\pi \sigma (d-1)/2},
\label{eq:h1}
\end{split}
\ee
with $4\mal{K}_d = \pi^{-(d+1)/2} \Gamma ((d-1)/2)$ and
$\sigma = \,{\rm sign} (t_1 - t_2)$. The corresponding expression for $m\neq 0$ depends on the analytic properties of the Bessel functions $K_\nu$ for vanishing argument and is quite more complicated. However, one can still identify the presence of a branching point for $\abs{t_1 - t_2} = r$ of the same nature as the one just discussed.
Taking twice the imaginary part of Eq.~\reff{eq:h1} yields Eq.~\reff{eq:RG} for the response function $R$. Note that this $R$ apparently vanishes for odd $d\neq 1$ because the exponent in $iG_{0E} (\tau_1 , \tau_2)$ is integer and thus $|t_1 - t_2| = r$ become simple poles instead of branching points. Of course, $R\equiv 0$ is unphysical and  the point is that  the limit in Eq.~\reff{eq:GE} has to be interpreted in terms of distributions: for example, in $d=3$ one would have $iG_0^< (t_1, t_2) \propto 1/[r^2 - (t_1-t_2 + i 0^+)^2]$, which corresponds to $P\frac{1}{r^2 - (t_1 - t_2)^2} + i\pi [\delta (r- t_1 + t_2)  - \delta (r + t_1 - t_2)]/(2r)$, where $P$ is the principal part. Thus the response function is actually $R(t,s) \propto -(\pi/r)\delta (r- t + s)$. This holds true also in higher odd dimensions, although the expression becomes progressively more complicated. 

\end{section}

\end{document}